\documentclass{iau} 
\usepackage{graphicx}
\usepackage{hyperref}

\title[The Gaia Archive] %% give here short title %%
{The Gaia Archive}

\author[Alcione Mora et al.]   %% give here short author list %%
{
Alcione Mora$^1$,
Juan Gonz\'alez-N\'u\~nez$^2$,
Deborah Baines$^2$,
Javier Dur\'an$^2$,
Ra\'ul Guti\'errez-Sanch\'ez$^2$,
Elena Racero$^2$,
Jes\'us Salgado$^2$
\and Juan Carlos Segovia$^2$
}

\affiliation{
$^1$ESA-ESAC Gaia Science Operations Centre, Camino Bajo del Castillo s/n, Urb. Villafranca del Castillo, 28692 Villanueva de la Ca\~nada, Madrid, Spain \\ email: {\tt alcione.mora@esa.int} \\[\affilskip]
$^2$ESA-ESAC Science Data Centre, Camino Bajo del Castillo s/n, Urb. Villafranca del Castillo, 28692 Villanueva de la Ca\~nada, Madrid, Spain}

\pubyear{2017}
\volume{330}  %% insert here IAU Symposium No.
\jname{Astrometry and Astrophysics in the Gaia Sky}
\editors{A. Recio-Blanco, P. de Laverny, A. Brown \& T. Prusti, eds.}
\begin{document}

\maketitle

\begin{abstract}
The Archive is the main Gaia data distribution hub. The contents of DR1 are briefly reviewed and the data structures discussed. The system architecture, based on Virtual Observatory standards, is also presented, together with the extensions that allow e.g. authenticated access, persistent uploads and table sharing. Finally some usage examples are provided.

\end{abstract}

\firstsection % if your document starts with a section,
\section{Introduction}

The main contents of Gaia Data Release 1 (DR1) are included in one big catalogue table: {\tt gaia\_source}, including astrometric parameters and average photometry for 1.14 billion sources. {\tt tgas\_source} contains the subsample of 2.06 million stars for which a good five parameters astrometric solution was obtained. Other data comprise a number of selected external catalogues, their corresponding pre-computed cross-matches and light curves for a sample of Cepheids and RR~Lyrae stars. Sect.~\ref{sect:contents} provides further details on the DR1 contents.

Astronomical archives are challenged by the need to provide increasingly larger data sets. The Gaia mission is surveying more than a billion sources. The final data release volume will be in the scale of petabytes. New projects and missions will push this boundary by several orders of magnitude (e.g. Euclid, LSST, SKA). Consequently, one key objective in modern archives is to carry out as much processing as possible on the server side, according to the ``code to the data'' paradigm. The Gaia Archive can be considered a step in this direction. Its architecture is described in Sect.~\ref{sect:architecture}, while some usage examples are shown in Sect.~\ref{sect:examples}.

\section{Archive contents}
\label{sect:contents}

For many science cases, DR1 can be considered just the big table {\tt gaia\_source}, providing astrometric and photometric information (columns) for each object (row). However, the Gaia Archive contains more information, which is presented below.

Light curves and detailed variability data for a sample of around 3000 pulsating variables (Cepheids and RR~Lyrae) in the LMC, due to the observing geometry during the initial phase of the nominal mission, when the ecliptic poles were repeatedly observed. The data are provided in a collection of flat tables. For example the {\tt rrlyrae} table includes a number of parameters for each RR~Lyrae (period, variability amplitude, Fourier decomposition, ...). The light curves are included in {\tt phot\_variable\_time\_series\_gfov}, which includes a row for each star and observing epoch. The latter structure is flat, simple and conveniente, but its applicability to the final Data Release is uncertain, because the table size would approach the trillion entries.

The Gaia data is most powerful when combined with additional information. A number of major surveys have been adapted and ingested into the Archive (2MASS, PPMXL, SDSS9 , UCAC4, URAT1, WISE). Pre-computed cross-matches have been generated for improved usability. For each catalogue, both the best neighbour and all possible matches in a given neighbourhood have been provided in dedicated pivot tables. The tables include one match per row, relating the Gaia to external catalogue IDs in different columns.

Finally, there is a number of additional catalogues which are considered immutable and are provided just for reference, such as Hipparcos, Tycho or IGSL. The latter is considered superseded by DR1 and should not be used for any future scientific analysis.

All tables considered, the total size of the Gaia Archive as of DR1 is 11.6 billion rows.

\section{Architecture}
\label{sect:architecture}

\begin{figure}
\begin{center}
\includegraphics[width=\hsize]{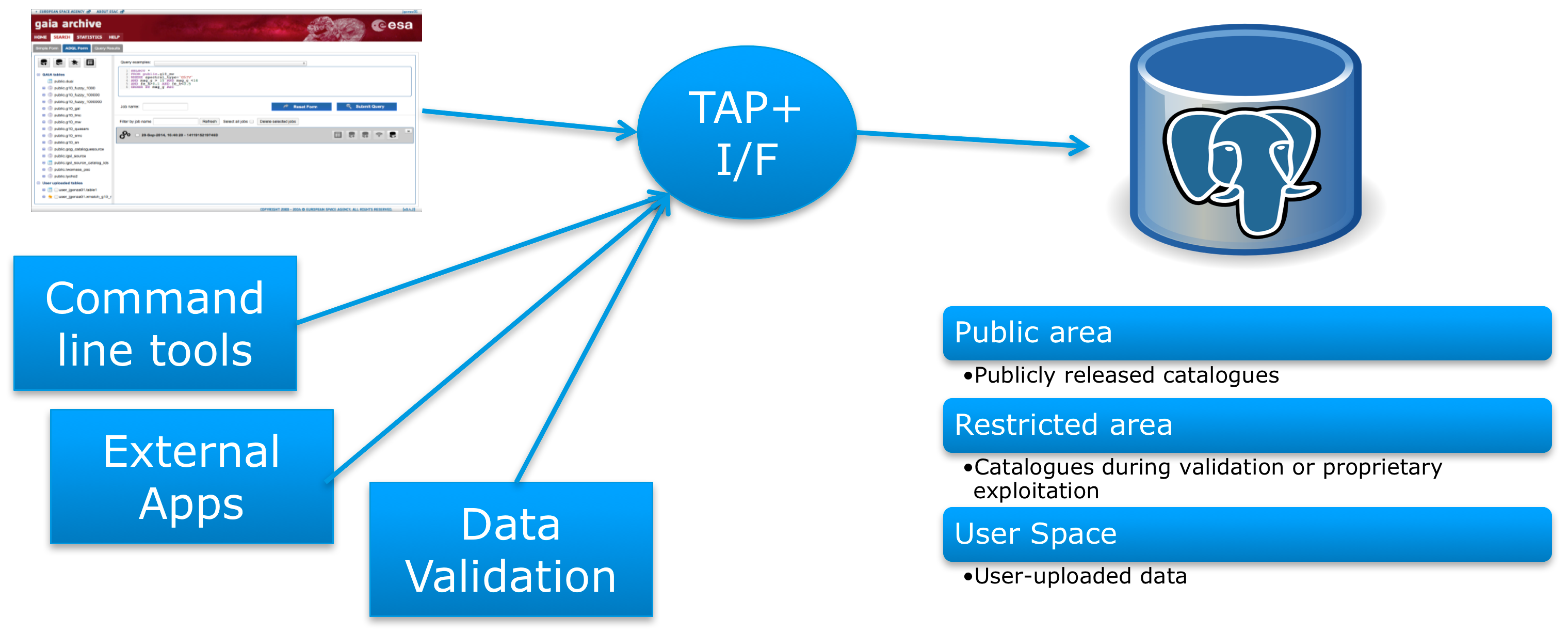}
\includegraphics[width=\hsize]{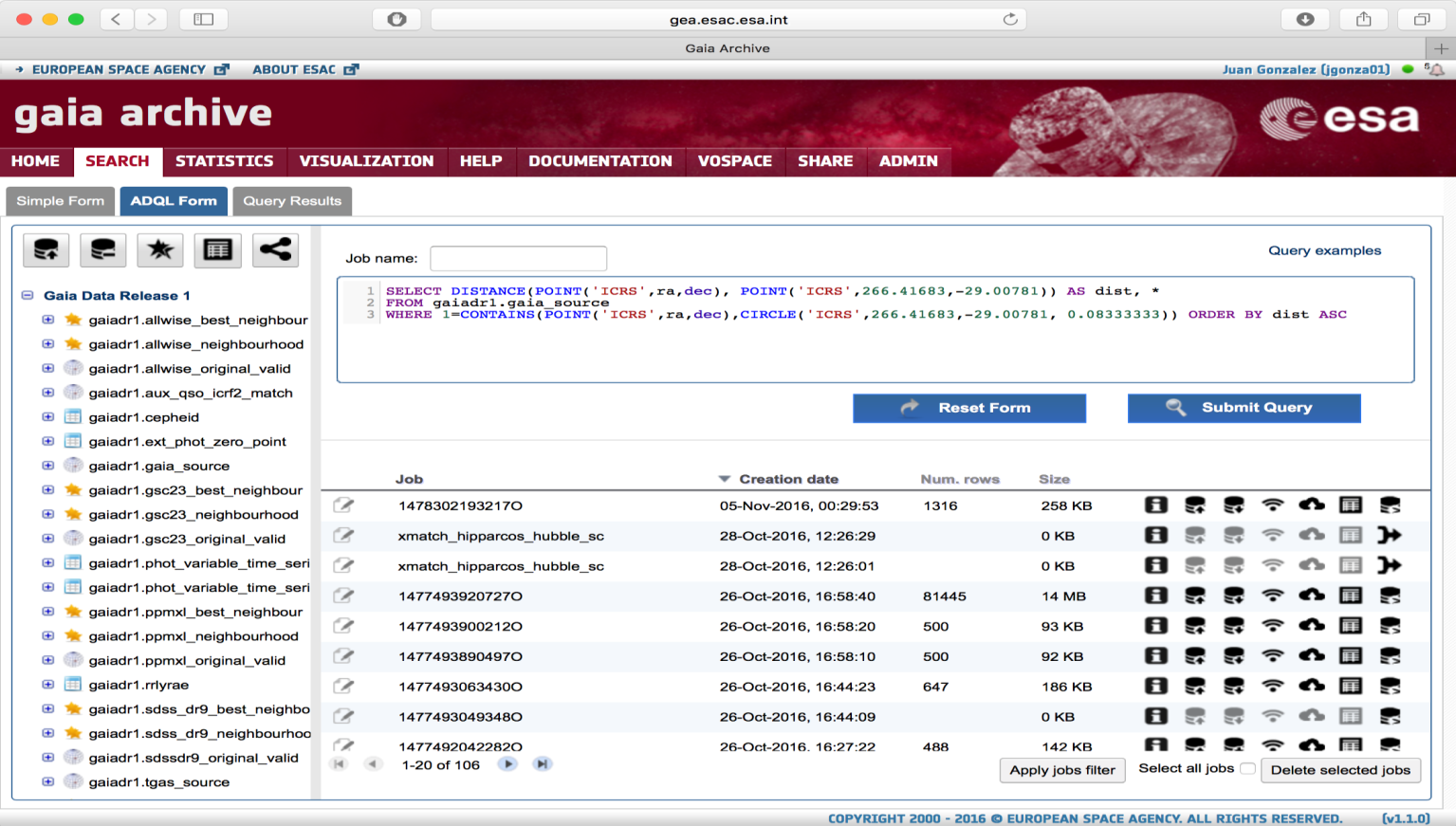}
\caption{Gaia Archive architecture (top) and Graphical User Interface (bottom)}
\label{fig:architecture}
\end{center}
\end{figure}

Fig.~\ref{fig:architecture} shows the main components of the Gaia Archive. The tables are stored in a PostgreSQL data base, and can be accessed through a VO TAP interface using the ADQL query language, a SQL dialect specifically designed for Astronomy. There are multiple ways to interact with the archive, including its own Graphical User Interface (GUI), external applications such as Topcat, command line tools and more recently a Python module within AstroPy.

Custom extensions on TAP, called TAP+, have been developed to provide additional functionalities outside the standard, most notably persistent uploads, table sharing and server cross-match. The key difference is that TAP is basically an anonymous protocol, while the extensions support user authentication to provide privacy. All jobs executed while logged in the system are associated to the user and kept until actively erased. User defined tables can be created either using the output of a job or uploading external data. Those tables can be used in queries exactly as if they were public data, but are only visible to their owner and those users with whom he decides to share them. This feature is very powerful. On the one hand, it provides the means to execute and refine complex queries in various steps. On the other hand, it allows the interactive analysis of preliminary results by a small number of people. Sharing is particularly useful during the validation campaigns before data releases. Query execution time and user space quotas can be defined on a per-user basis. Finally, an optimised geometrical cross-match service is offered, which allows users to carry out cone searches between tables with sky coordinate information.

The GUI can be accessed using any browser at the Archive address$\footnote{\url{http://archives.esac.esa.int/gaia}}$. It offers the possibility to carry out basic and advanced queries and to quickly inspect the output. The data can be downloaded in a variety of formats, including VO table, fits and csv or directly exported to other VO tools via SAMP. Help is provided in different ways. A dedicated tab in the GUI contains general information, tutorials, explanations on ADQL custom functions, and instructions for command line and programmatic access. Information on each table and column is available from a tree view. Units are shown in the table preview and exported into output files. Links to external resources are also provided (e.g. DR1 documentation and data model detailed description).

\section{Examples}
\label{sect:examples}

The use of ADQL to query the Gaia Archive has pros and cons. The two biggest advantages are probably flexibility and reproducibility. Complex operations can be defined within a query to retrieve derived quantities (e.g. absolute magnitudes, colours, variability phase, ...). Bute even more important is the ability to exactly reproduce the results of any previous work where the original ADQL queries are included.

\begin{figure}
\begin{center}
\begin{minipage}[b]{0.55\hsize}
\large
\begin{verbatim}
SELECT * 
FROM gaiadr1.tgas_source  
WHERE CONTAINS(
    POINT('ICRS', ra, dec),
    CIRCLE('ICRS', 56.75, 24.1167, 2)
    )=1 
AND pmra IS NOT NULL AND pmra != 0 
AND pmdec IS NOT NULL AND pmdec != 0 
AND abs(pmra_error/pmra) < 0.10 
AND abs(pmdec_error/pmdec) < 0.10 
AND pmra BETWEEN 15 AND 25
AND pmdec BETWEEN -55 AND -40 
\end{verbatim}
\end{minipage}
\includegraphics[width=0.44\hsize]{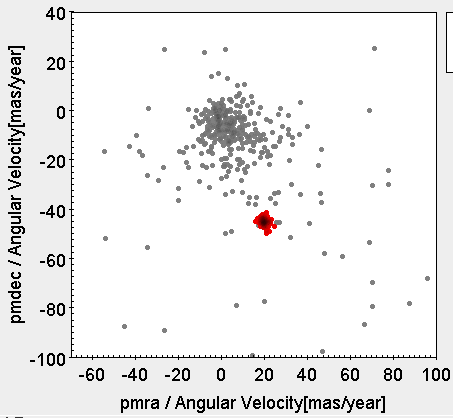}
\caption{Simple selection of Pleiades member candidates. Left: a simple ADQL query shows how to explore a circular region on the sky adding extra constraints on the proper motion. Right: the candidate members (red) show a clearly distinct behaviour compared to field stars in proper motion space.}
\label{fig:m45Cluster}
\end{center}
\end{figure}

Being a relatively new tool, learning ADQL can be an entrance barrier to advanced Archive users. The learning curve is relatively shallow, though, due to the reduced syntax and simplicity of the language. Fig.~\ref{fig:m45Cluster} shows a simple query used to retrieve candidate members of the Pleiades open cluster based on position on the sky (two degrees radius circle) and proper motion (box of size 20 mas/yr in RA and 15 mas/yr in DEC approximately centred on the average value). The selected objects in this region can be clearly separated from field stars in proper motion space.

\begin{figure}
\begin{center}
\includegraphics[height=4cm]{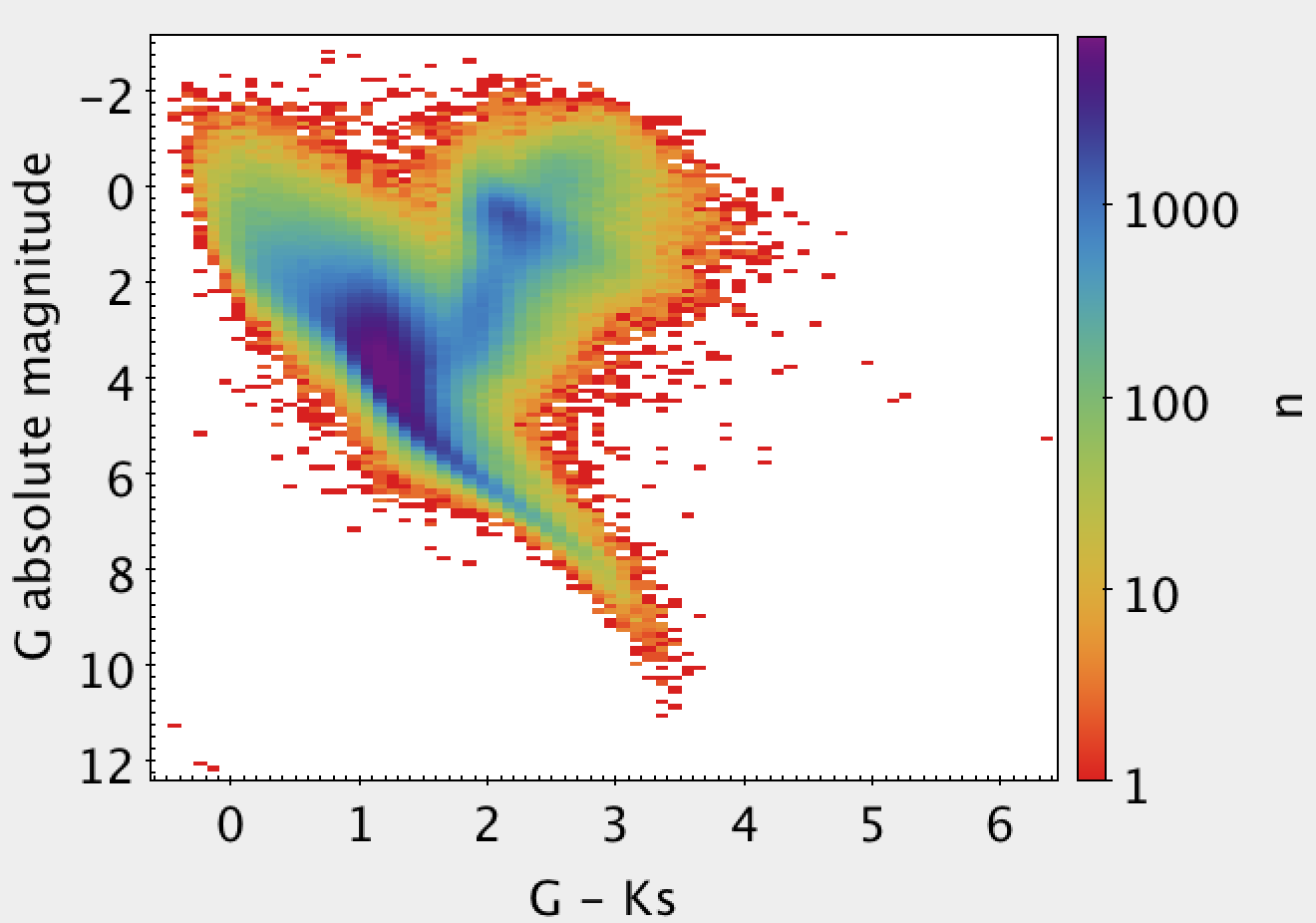}
\includegraphics[height=4cm]{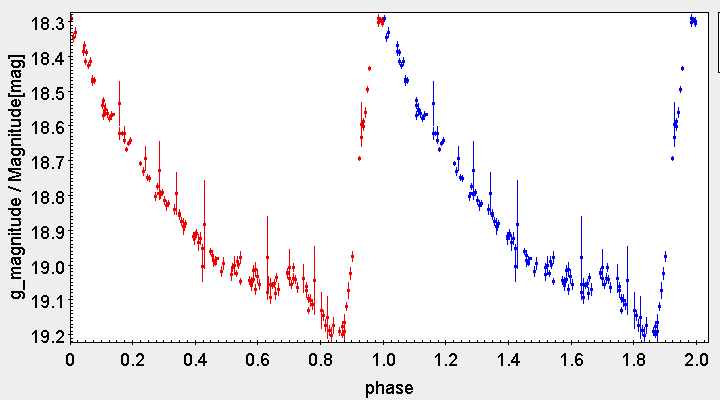}
\caption{Example DR1 plots based on \cite[Gaia Collaboration et al. (2016)]{2016A&A...595A...2G}. Left: TGAS HR diagram computed as a 2D histogram. Right: phase folded light curve for RR Lyrae with source\_id 5284240582308398080.}
\label{fig:examplesBrown2016}
\end{center}
\end{figure}

\cite[Gaia Collaboration et al. (2016)]{2016A&A...595A...2G}. provides several examples of ADQL queries illustrating the contents of DR1. Fig.~\ref{fig:examplesBrown2016} adapts two of them: the TGAS HR diagram and the phase folded light curve of an RR~Lyrae star. The original HR diagram was a scatter plot containing more than a million $G-Ks$ colours and $M_G$ absolute magnitudes. While this is OK for DR1, representing a billion points in a diagram for DR2 would be very difficult and most probably an overkill. The figure shows how to provide the same information using a memory efficient 2D histogram. The RR Lyrae light curve has been folded, and the errors have been directly estimated using an ADQL query within the Archive.

\section*{Acknowledgements}
This work has made use of data from the European Space Agency (ESA)
mission {\it Gaia} (\url{https://www.cosmos.esa.int/gaia}), processed by
the {\it Gaia} Data Processing and Analysis Consortium (DPAC,
\url{https://www.cosmos.esa.int/web/gaia/dpac/consortium}). Funding for
the DPAC has been provided by national institutions, in particular the
institutions participating in the {\it Gaia} Multilateral Agreement.

\end{document}